\documentclass[prb,twocolumn,showpacs,preprintnumbers,amsmath,amssymb]{revtex4-1}
\usepackage{graphicx}

\begin{document}

\title{Electronic structure of BaNi$_2$P$_2$ observed by angle-resolved photoemission spectroscopy}

\author{S.~Ideta$^{1}$, T.~Yoshida$^{1}$, M.~Nakajima$^{1, 2, 6}$, W.~Malaeb$^{3}$, H.~Kito$^{2,6}$, H.~Eisaki$^{2, 6}$, A.~Iyo$^{2, 6}$, Y.~Tomioka$^{2, 6}$, T.~Ito$^{2, 6}$, K.~Kihou$^{2, 6}$, C.~H.~Lee$^{2, 6}$, Y.~Kotani$^{4}$, K.~Ono$^{4}$, S.~K.~Mo$^7$, Z.~Hussain$^7$, Z.-X.~Shen$^8$, H.~Harima$^{5, 6}$, S.~Uchida$^{1, 6}$, and A.~Fujimori$^{1, 6}$}

\affiliation{
$^{1}$Department of Physics, University of Tokyo, Bunkyo-ku, Tokyo 113-0033, Japan\\
$^{2}$National Institute of Advanced Industrial Science and Technology (AIST), Tsukuba, Ibaraki 305-8568\\
$^{3}$ Institute for Solid State Physics (ISSP), University of Tokyo, Kashiwa-no-ha, Kashiwa, Chiba 277-8581, Japan\\
$^{4}$Photon Factory, Institute of Materials Structure Science, KEK, Tsukuba, Ibaraki 305-0801, Japan \\
$^{5}$Department of Physics, Graduate School of Science, Kobe University, Kobe 657-8501\\
$^{6}$JST, Transformative Research-Project on Iron Pnictides (TRIP), Chiyoda, Tokyo 102-0075\\
$^{7}$Advanced Light Source, Lawrence Berkeley National Lab, Berkeley, California 94720, USA\\
$^{8}$Department of Physics, Applied Physics, and Stanford Synchrotron Radiation Laboratory, Stanford University, Stanford, California 94305, USA
}
\date{\today}%
\begin{abstract}
We have performed an angle-resolved photoemission spectroscopy (ARPES) study of BaNi$_2$P$_2$ which shows a superconducting transition at $T_c$ $\sim$ 2.5 K. We observed hole and electron Fermi surfaces (FSs) around the Brillouin zone center and corner, respectively, and the shapes of the hole FSs dramatically changed with photon energy, indicating strong three-dimensionality. The observed FSs are consistent with band-structure calculation and de Haas-van Alphen measurements. The mass enhancement factors estimated in the normal state were $m^*$/$m_b$ $\leq$ 2, indicating weak electron correlation compared to typical iron-pnictide superconductors. An electron-like Fermi surface around the Z point was observed in contrast with BaNi$_2$As$_2$ and may be related to the higher $T_c$ of BaNi$_2$P$_2$.
\end{abstract}

\maketitle
\section{Introduction}
For more than two decades, ternary pnictides $A$Ni$_2$$\rm{Pn}_2$, where $A$ and $\rm{Pn}$ are an alkali-earth atom and a pnictogen atom, respectively, have been known to show a variety of magnetic properties such as Pauli paramagnetism (e.g., LaNi$_2$P$_2$), anti-ferromagnetism (e.g., GdNi$_2$P$_2$), and ferromagnetism (e.g., PrNi$_2$P$_2$, NdNi$_2$P$_2$) \cite{Jeitschko}. Recently, one of the $A$Ni$_2\rm{Pn}_2$ materials, BaNi$_2$P$_2$, was found to show superconductivity below $\sim$2.5 K \cite{Mine_BaNiP,Tomioka}. Other Ni compounds, BaNi$_2$As$_2$ \cite{Ronning}, SrNi$_2$As$_2$ \cite{Bauer}, and SrNi$_2$P$_2$ \cite{Ronning_2} also show lower superconducting transition temperatures ($T_c$'s) of 0.6 - 1.4 K. Although these $T_c$'s are much lower than the $T_c$'s of the iron pnictide superconductors (Fe-SCs) with the same crystal structure such as Ba$_{1-x}$K$_x$Fe$_2$As$_2$ and Ba(Fe$_{1-x}$Co$_x$)$_2$As$_2$ with $T_c$'s exceeding 20 K \cite{Sefat, Rotter}, studies of the electronic structures of the low-$T_c$ materials and their comparison with those of the high-$T_c$ materials are expected to give an important clue to understand the origin of the high-$T_c$ superconductivity in the iron pnictides. 

BaNi$_2$As$_2$, which has $T_c \sim$ 0.7 K and shows a first-order-like structural transition from a tetragonal to a triclinic structure at $T_s \sim$ 130 K without magnetic order, has been studied by angle-resolved photoemission spectroscopy (ARPES) \cite{BZhou}. The electronic structure near the Fermi level ($E_F$) has been shown to change with the structural phase transition, and  part of the Fermi surfaces (FSs) disappears below $T_s$. BaNi$_2$P$_2$ does not show magnetic nor structural transition, yet shows the highest $T_c \sim$ 2.5 K among the $A$Ni$_2\rm{Pn}_2$ family. Furthermore, $A$Ni$_2$P$_2$ ($A$ = Ba, Sr) shows higher $T_c$'s than $A$Ni$_2$As$_2$: the $T_c$ decreases in the order BaNi$_2$P$_2$ $>$ SrNi$_2$P$_2$ $>$ BaNi$_2$As$_2$ $\geq$ SrNi$_2$As$_2$. Therefore, it is intriguing to clarify the difference in the electronic structure between BaNi$_2$P$_2$ and BaNi$_2$As$_2$ and to identify which aspect contributes to the differences in the $T_c$.

Recently, a de Haas-van Alphen (dHvA) study of BaNi$_2$P$_2$ has revealed the shapes of the FSs with strong three-dimensionality, consistent with the band-structure calculation, and the estimated mass enhancement factors are in the range of $\sim$ 2 - 3 \cite{Terashima}. In this work, we have performed an ARPES study of BaNi$_2$P$_2$ with $T_c$ = 2.5 K studied by using energy tunable, polarized photons from synchrotron radiation. The result was found to be in good agreement with the band-structure calculation, but the band dispersion was found to be renormalized by factor of  1 - 2 for the hole and electron bands, indicating weaker electron correlation than the Fe-SCs. A large electron-like FS around the Z point [$k$=(0 ,0, 2$\pi/c$)] was observed in contrast to BaNi$_2$As$_2$, where only a small hole pocket was observed around the Z point below the structural transition temperature. This implies that the existence of the electron FS around the Z point may contribute to the higher $T_c$ of BaNi$_2$P$_2$. The electron-like FS is absent around the $\Gamma$ point in BaNi$_2$P$_2$, indicating that the strong FS warping occurs in the $k_z$ direction, that is, the FS around the BZ center shows strong three-dimensionality. We suggest that not only the three-dimensionality of the FSs but also the intralayer $\rm{Pn-Pn}$ distance or the pnictogen height may be an important parameters to determine the $T_c$ in the $A$Ni$_2$Pn$_2$ family.

\begin{figure}[t]
\begin{center}
\includegraphics[width=7cm]{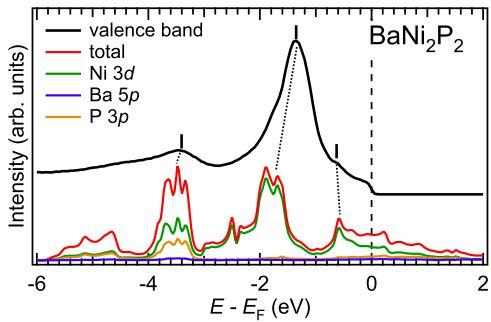}
\end{center}
\caption{(Color online) Angle-integrated photoemission spectrum of BaNi$_2$P$_2$ in the valence-band region measured with $h\nu$ = 85 eV (black curve) and the total and partial electron density of states calculated using a WIEN2k program.}
\label{Fig1}
\end{figure}

\section{Experiment}

Single crystals of BaNi$_2$P$_2$ ($T_c$ $\sim$ 2.5 K) were prepared by a high-pressure synthesis method using a cubic-anvil type apparatus. ARPES experiments were carried out at beamline 28A of Photon Factory (PF) and beamline 10.0.0.1 of Advanced Light Source (ALS). Measurements were performed at $T$ = 10 K. Photon energy was set at $h\nu$ = 38 - 88 eV with circularly polarized light at PF and linearly polarized light at ALS. Samples were cleaved $in$-$situ$. The total energy resolution was set at $\Delta E$ $\sim$20 - 30 meV. The FSs of BaNi$_2$P$_2$ were calculated within the local density approximation by using the full potential LAPW (FLAPW) method. We used the program codes TSPACE \cite{Yanase} and KANSAI-06. We also used WIEN2k package \cite{WIEN2k} in order to get the band structure and orbital characters. 

\section{results and discussion}

\begin{figure}[t]
\begin{center}
\includegraphics[width=9.2cm]{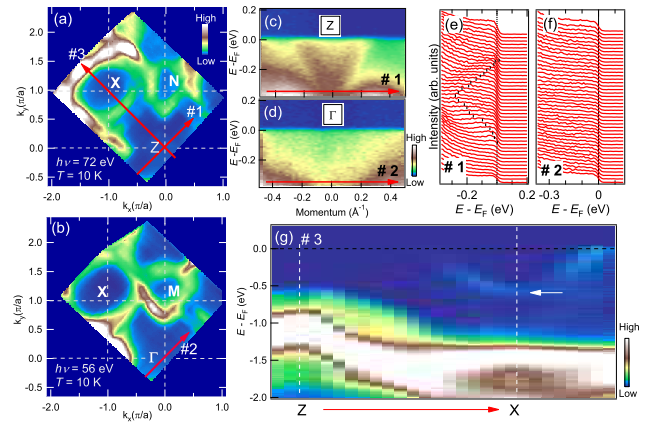}
\end{center}
\caption{(Color online) ARPES spectra of BaNi$_2$P$_2$ taken at $T$ = 10 K with  $h\nu$ = 72 and 56 eV (circularly polarized light) on the $k_x$ - $k_y$ planes containing the Z and $\Gamma$ points, respectively. (a), (b): Fermi surface mapping integrated within $\pm$ 20 meV of $E_F$. (c), (d): Band dispersions for cuts $\#$1 and $\#$2 in panels (a) and (b). (e), (f):  Energy distribution curves (EDCs) corresponding to panels (c) and (d). Black bars are guide to the eye. (g): Band dispersions for cut $\#$3 in (a) along the Z - X direction. A white arrow shows the minimum of the electron band. Measurements were performed using circularly polarized light.}
\label{Fig2}
\end{figure}

In Fig. \ref{Fig1}, the valence-band spectrum of BaNi$_2$P$_2$  taken at $h\nu$ = 85 eV is compared with the electronic density of state (DOS) given by the band-structure calculation. The calculated total DOS of BaNi$_2$P$_2$ is dominated by the Ni 3$d$ states near the $E_F$. The angle-integrated spectrum well agrees with the calculated DOS\cite{Shein}, which shows that all the Ni-based 122 compounds such as BaNi$_2$P$_2$, BaNi$_2$As$_2$, SrNi$_2$As$_2$, and SrNi$_2$P$_2$ are metallic. The small energy difference between the main peak at -1.5 eV and that of the band-structure calculation is probably due to a weak mass renormalization caused by electron correlation.

Figures 2(a) and (b) show FS mapping where the ARPES intensity has been integrated within $\pm$ 20 meV of $E_F$. They were taken at $h\nu$ = 72 and 56 eV corresponding to the out-of plane momenta $k_z$ of the $\sim$ Z [$k_z \sim$ 9 (2$\pi/c$)] and $\sim \Gamma$ [$k_z \sim$ 8.1 (2$\pi/c$)] points, respectively. Here, we determined the $k_z$ value from photon-energy dependence experiment and the inner potential was set at 17 eV. The ARPES intensity plotted in energy-momentum ($E$-$k$) space near the Z and $\Gamma$ points for cuts $\#$1 and $\#2$ are shown in Figs. \ref{Fig2}(c) and \ref{Fig2}(d), respectively. Figure \ref{Fig2}(e) shows the EDCs corresponding to the intensity plot in Figs. \ref{Fig2}(c). The spectra near the Z point show a parabolic electron-like band dispersion while such a band is absent around the $\Gamma$ point [cut $\#$2 shown in Figs. \ref{Fig2}(d) and \ref{Fig2}(f)]. This difference comes from the three dimensionality of the electron FS around the BZ center. Band dispersions along the Z - X line (cut $\#$3) are shown in Fig. \ref{Fig2}(g). Another electron band with a larger FS volume is observed around the X point with the bottom as deep as $\sim$ - 0.6 eV indicated by a white arrow, which is nearly the same value as that in BaNi$_2$As$_2$ \cite{BZhou}.

\begin{figure}[t]
\begin{center}
\includegraphics[width=8.5cm]{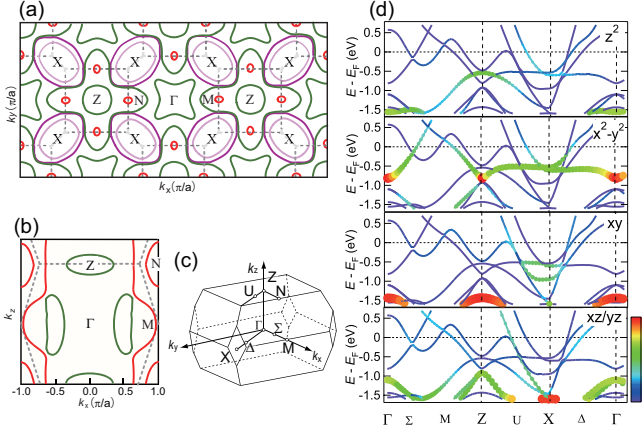}
\end{center}
\caption{(Color online) Fermi surface (FS) cross-sections in the $k_x$ - $k_y$ and $k_z$ - $k_x$ planes of BaNi$_2$P$_2$ given by the band-structure calculation. (a): $k_x$ - $k_y$ cuts at the Z and the $\Gamma$ points. (b): $k_z$ - $k_x$ plane cuts containing the $\Gamma$ and Z points. Gray dotted lines show the Brillouin zone boundaries. (c): Three-dimensional Brillouin zone. (d): Results of the band-structure calculation with orbital character. The contribution of the orbitals is represented by both the size of the symbols and the color scale.}
\label{Fig3}
\end{figure}

\begin{figure}[t]
\begin{center}
\includegraphics[width=8.8cm]{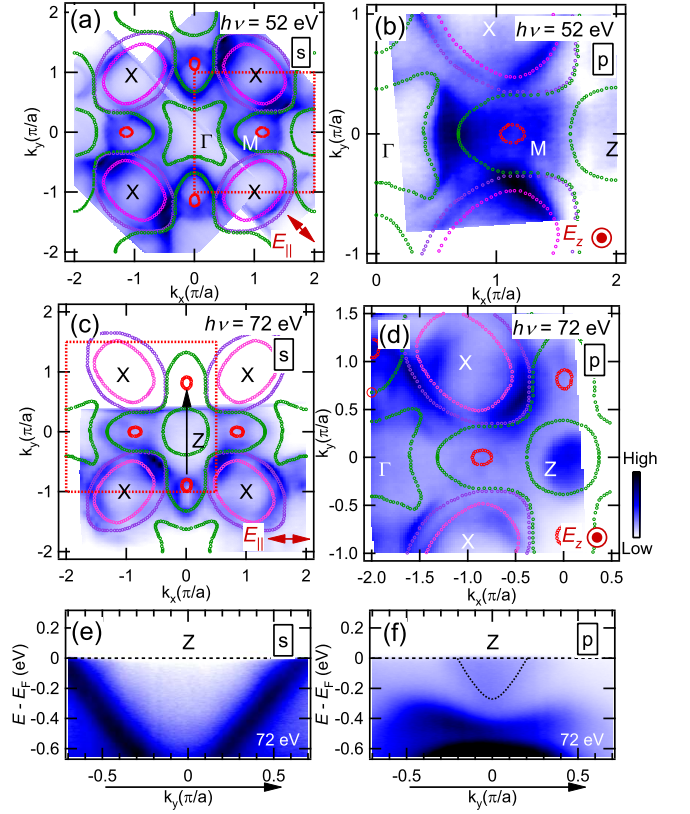}
\end{center}
\caption{ (Color online) ARPES intensity map measured with $s$ and $p$ polarizations and several photon energies. (a),(b): FS mapping taken at $h\nu$ = 52 eV with the $s$ and $p$ polarizations, respectively, corresponding to the $k_x$-$k_y$ plane containing the $\Gamma$ point. Comparison with the calculated FSs. (b)  Magnified view of the area enclosed by a red square in (a). (c),(d): FS mapping taken at $h\nu$ = 72 eV with the $s$ and $p$ polarizations, respectively, corresponding to the $k_x$-$k_y$ plane containing the Z point. (d) is a magnified view of the area enclosed by a red square in (c). (e), (f): Band dispersions along the Z - $\Gamma$ line shown as an arrow in (c) for the $s$ and $p$ polarizations, respectively. A dotted curve in (f) is guide to the eye. All data were measured at $T$ = 10 K, and polarization vectors labeled in red.}
\label{Fig4}
\end{figure}

\begin{figure}[t]
\begin{center}
\includegraphics[width=9cm]{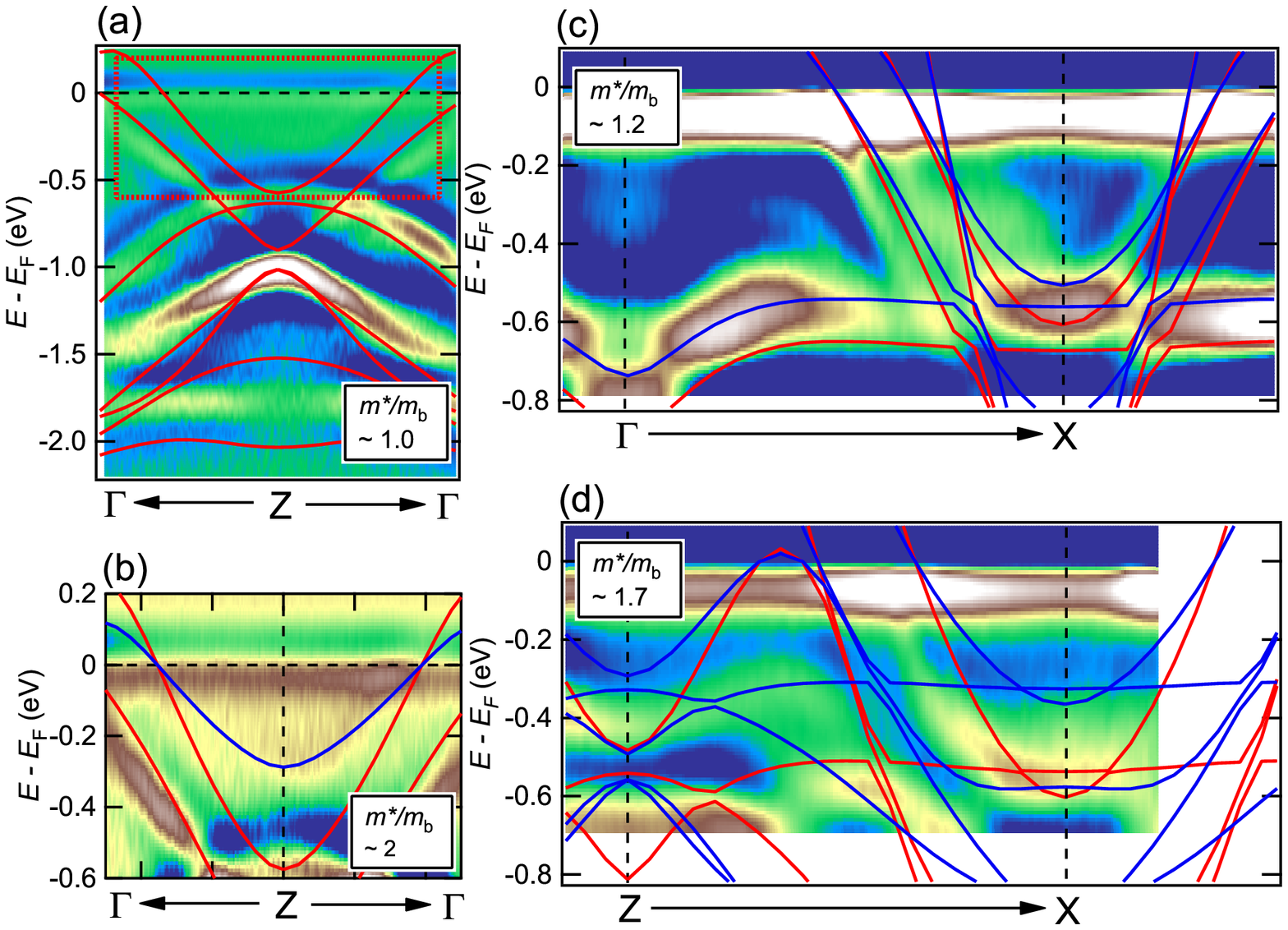}
\end{center}
\caption{(Color online) Comparison of the second-derivative $E$ - $k$ plots of ARPES data with the band-structure calculation. Red and blue curves show results of the band-structure calculation and renormalized bands ($m^*/m_b$ $>$ 1), respectively. Here, $m^*$ and $m_b$ denote the band masses determined by experiment and by the band-structure calculation, respectively. (a): Hole bands around the Z point taken at $h\nu$ = 74 eV with circularly polarized light. (b): Magnified plot near $E_F$ around the Z point. (c),(d): Band dispersions in the wide momentum range taken at $h\nu$ = 52 eV and 72 eV, corresponding to the $\Gamma$ - X and Z - X lines, respectively.}
\label{Fig5}
\end{figure}

Calculated FSs in the $k_x$ - $k_y$ plane around the Z and $\Gamma$ points are shown in Fig. 3(a) and those in the $k_z$ - $k_x$ plane in Fig. \ref{Fig3}(b). A small hole pocket elongated along the $k_z$ direction (red) encloses the N and M points. A large cylindrical FS centered at the BZ corner (purple and pink) is electron-like. The other FSs around the Z point indicated by green curves are electron-like. As shown in Fig. \ref{Fig3}(c), the FSs shown by green curves in the $k_z$ - $k_x$ plane show strong three-dimensionality as mentioned above. On the other hand, the FSs at the BZ corner have relatively weak three-dimensionality (not shown here) \cite{Terashima}. Figure \ref{Fig3}(d) shows band dispersions and thier orbital character. One can see that some bands (e.g., the electron-like band at the Z point) have little Fe 3$d$ orbital character (as in the case of BaNi$_2$As$_2$ \cite{BZhou}), indicating that they are dominated by P 3$p$  (As 4$p$) orbital for BaNi$_2$P$_2$ (BaNi$_2$As$_2$). The stronger $p$ orbital character than that in the Fe pnictides can be attributed to the fact that the Ni 3$d$ levels are located at lower energies than the Fe 3$d$ levels.

\begin{figure}[t]
\begin{center}
\includegraphics[width=7.5cm]{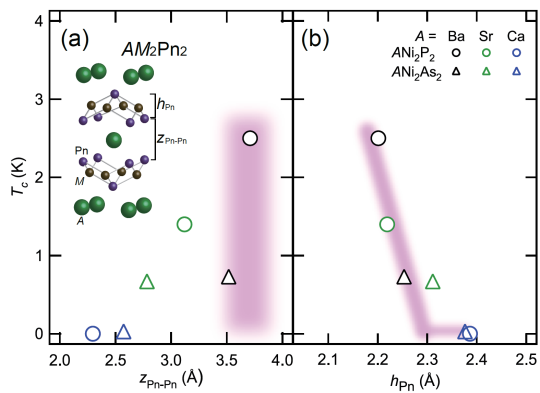}
\end{center}
\caption{(Color online) (a),(b): Relationship between $T_c$, the inter-plane $\rm{Pn}$-$\rm{Pn}$ distance $z_{\rm{Pn}-\rm{Pn}}$, and the pnictogen height ($h_{\rm{Pn}}$) for $A$Ni$_2$\rm{Pn}$_2$ compounds, where $A$ = Ba, Sr, Ca;  $\rm{Pn}$ = P, As. Definition of $z_{\rm{Pn}-\rm{Pn}}$ and $h_{\rm{Pn}}$ is shown in the inset of panel (a). Data for the $z_{\rm{Pn}-\rm{Pn}}$ and the pnictogen height $h_{\rm{Pn}}$ are taken from Ref. [16].}
\label{Fig6}
\end{figure}

The ARPES-intensity map in the $k_x$ - $k_y$ plane is compared with the calculated FSs in Figs. \ref{Fig4}(a) - (d). The FSs around the Z and $\Gamma$ points well accord with the band-structure calculation: the large electron FSs around the BZ corner taken at $h\nu$ = 72 and 52 eV [$k_z$ $\sim$ 8.9 and $\sim$ 8 (2$\pi/c$), respectively] are in good agreement with the band-structure calculation. By using the $s$ and $p$ polarizations \cite{YZhang, Nishi}, the calculated parabolic band at the BZ center (Z) shown in Fig. \ref{Fig3}(d) was observed in the $p$ polarization experiment in Fig. \ref{Fig4}(d), while the band disappeared for the $s$ polarization, indicating that the parabolic band is attributed to the $d_{xy}$ orbital as in the previous ARPES study on BaNi$_2$As$_2$ \cite{BZhou}. On the other hand, in Fig. \ref{Fig4}(b), even by using the same $p$ polarization, the FS at the BZ center (Z) is not observed \cite{fot1}. As shown in Fig. \ref{Fig4}(e), the parabolic band was not observed in the $s$ polarization measurement, while in Fig. \ref{Fig3}(d) the predominantly contributing $d_{x^2-y^2}$ band was clearly observed for the same polarization. For the $p$ polarization, the parabolic band near $E_F$ and a band with $d_{z^2}$ orbital character at -0.6 eV below $E_F$ were clearly observed as shown in Fig. \ref{Fig4}(f), in good agreement with the band-structure calculation.

Next, we discuss the band renormalization for the hole and electron bands. In Fig. \ref{Fig5}, the hole and electron bands along high symmetry lines are shown. We have estimated the mass renormalization factor ($m^*/m_b$) by comparing the experimental data with the band-structure calculation shown by red and blue curves corresponding to the bare bands and renormalized bands, respectively. The second-derivative $E$ - $k$ plots shown in Fig. \ref{Fig5}(a) well agree with the band-structure calculation shown by red curves, namely, $m^*/m_b \sim$ 1, while the parabolic band near $E_F$ around the BZ center shown in Fig. \ref{Fig5}(b) is strongly renormalized by a factor of $\sim$ 2 as shown by a blue curve. This $m^*/m_b$ value is a little smaller than the result of the dHvA study, $m^*/m_b$ $\sim$ 2.8 - 3.1 \cite{Terashima}. In Figs. \ref{Fig5}(c) and \ref{Fig5}(d), the second-derivative $E$ - $k$ plots in the wide momentum range are shown and the calculated band structure (red) and the renormalized one (blue) are superimposed. One finds that the energy minima of the electron bands around the X point are renormalized by a factor of $m^*/m_b$ $\sim$ 1.6 - 1.7, which is consistent with the dHvA study \cite{Terashima}, and the other bands are almost $m^*/m_b$ $\sim$ 1. The fact that the mass enhancement factor experimentally deduced in the present ARPES study is smaller than that deduced from the dHvA study may indicate that additional mass renormalization occurs in the vicinity of $E_F$.

Finally, we discuss which parameters predominately influence the $T_c$ among the ternary Ni pnictides. The $T_c$'s of $A$Ni$_2$Pn$_2$ are plotted as functions of the interlayer Pn-Pn distance ($z_{\rm{Pn-Pn}}$) and the pnictogen height ($h_{\rm{Pn}}$) in Fig. \ref{Fig6}. Figure \ref{Fig6}(a) shows that the $T_c$ generally increases with increasing $z_{\rm{Pn-Pn}}$, which controls the three dimensionality of FSs (although the $z_{\rm{Pn-Pn}}$ dependence shows some irregularity). On the other hand, the $T_c$ depends on $h_{\rm{Pn}}$ more sensitively and systematically as shown in Fig. \ref{Fig6}(b). Therefore, the lower $h_{\rm{Pn}}$ is and the longer $z_{\rm{Pn-Pn}}$ is, the higher the $T_c$ will be in the ternary Ni pnictides. Because the $h_{\rm{Pn}}$ controls the degeneracy of the Ni 3$d$ orbitals, it controls the topology and shapes of FSs in a complicated manner. The $z_{\rm{Pn-Pn}}$ controls the degree of three dimensionality, and hence the degree of FS nesting as well as the FS topology \cite{Bianconi}. In fact, the most prominent differences in the FS topology between the superconducting BaNi$_2$P$_2$ and the non-superconducting BaNi$_2$As$_2$ is the presence of an electron-like FS around the Z point in BaNi$_2$P$_2$.


\section{CONCLUSIONS}

We have performed ARPES experiments on BaNi$_2$P$_2$ with $T_c$ $\sim$ 2.5 K. The Fermi surface shapes are consistent with the band-structure calculation. The Fermi surfaces show strong three-dimensionality, and the FS nesting should be weak. The mass renormalization factors are comparable or somewhat smaller than those estimated by the dHvA experiment, and are in the range of 1 $<$ $m^*/m_b$ $<$ 2: the electron-like FS which has been observed around the Z point shows a relatively strong mass enhancement of $\sim$ 2. This classifies BaNi$_2$P$_2$ as a weakly correlated superconductor. The electron-like Fermi surface around the Z point, which is absent in BaNi$_2$As$_2$ at low temperatures, may be related with the higher $T_c$ of BaNi$_2$P$_2$. 

\section*{Acknowledements}
The authors acknowledge S. Ishida for informative discussions. ARPES experiments were carried out at KEK-PF (Proposals No. 2009S2-005) and ALS (Proposal No. ALS-05054). This work was supported by an A3 Foresight Program from JSPS and, a Grant-in-Aid for Scientific Research on Innovative Area \textquotedblleft Materials Design through Computics: Complex Correlation and Non-Equilibrium Dynamics\textquotedblright, MEXT, Japan. SI acknowledges support from the Japan Society for the Promotion of Science for Young Scientists.


\begin{thebibliography}{9}
\bibitem{fot1} The $s$ polarization component increased its weight due to the geometry between the electron analyzer and the electric field vector of synchrotron radiation, namely, because the mirror plane was not perpendicular to the sample surface.
\end{thebibliography}
\end{document}